\documentclass[12pt]{article}

\advance\textwidth by 0.5in
\advance\hoffset by -0.25in

\usepackage{epsf}

\def\gsim{ \,\, \vcenter{\hbox{$\buildrel{\displaystyle >}\over\sim$}}
 \,\,}
\newcommand{\be}{\begin{eqnarray}}
\newcommand{\ee}{\end{eqnarray}}
\newcommand{\non}{\nonumber\\}
 \newcommand{\ave}[1]{\left\langle #1 \right\rangle}
 \newcommand{\absol}[1]{\left| #1 \right|}
 \newcommand{\GeV}{\,\hbox{GeV}}
 \newcommand{\MeV}{\,\hbox{MeV}}
 \newcommand{\inline}[1]{\noalign{\hbox{#1}}}

\begin{document}

\thispagestyle{empty}
\title {\bf Probing Gluons in Nuclei: the Case of $\eta'$}

\author
{
 Jamal Jalilian-Marian$^1$ and Sangyong Jeon$^{2,3}$
 \\
 {\small\it $^1$Physics Department, BNL, Upton, NY 11973, USA}\\
 {\small\it $^2$Department of Physics, McGill University, 
   Montreal, QC H3A-2T8, Canada}\\
 {\small\it and}\\
 {\small\it $^3$RIKEN-BNL Research Center, BNL, Upton, NY 11973, USA}\\
}

 \maketitle

 \begin{abstract}
Using the recently proposed $gg\eta'$ effective vertex, 
we investigate the production of $\eta'$ from gluon fusion in $pA$ collisions.
We show that measuring $\eta'$ production cross-section at moderate 
$x_{\eta'}$
yields direct information on the small $x$ gluon distribution function of 
the nucleus.  At RHIC, the smallest accessible $x$ turns out to be
$O(10^{-5})$ and at LHC, it is $O(10^{-8})$.  
Therefore, $\eta'$ is an excellent probe of the Color Glass Condensate.
\end{abstract}

 \newpage

 \section{Introduction}

 The $\eta'$ meson has many interesting properties, the most discussed being
 its unusual mass. 
 By spontaneously breaking $U(3)$ symmetry of the hadronic Lagrangian,
 one would expect that $\eta'$ would be a Goldstone boson whose finite
 mass is due to the finite mass of the quarks. 
 In reality, the $\eta'$ mass $(M = 0.958\GeV)$ is much heavier
 than the mass of the $\eta$ $(M_\eta = 0.547\GeV)$.
 The quark model
 Gell-Mann-Okubo formula for the singlet and the octet mass
 \be
 M_0^2 & = & 
 {f_\pi^2\over 3 f_0^2}
 \left(
 m_{K^+}^2
 +
 m_{K^0}^2
 +
 m_{\pi^+}^2
 \right)
 \\
 \inline{ and }
 M_8^2 
 & = & 
 {1\over 3}
 \left(
 2 m_{K^+}^2
 +
 2m_{K^0}^2
 +
 2 m_{\pi^+}^2
 +
 m_{\pi^0}^2
 \right)
 \ee
 is incapable of explaining this large mass difference between the
 (mostly) flavor octet state $\eta$ and the (mostly) singlet state
 $\eta'$.  In fact, the singlet mass $M_0$ is smaller than the octet
 mass $M_8$.
 Therefore, in the zero quark mass limit, about half of the $\eta'$ mass
 remains while the octet pseudo-scalar meson ($\pi, K, \eta$) masses
 go to zero. 

 The resolution of this $U_A(1)$ problem was provided by 
 't~Hooft \cite{'tHooft:1976up} and Witten \cite{Witten:1979vv}.
 In these works, it was argued that
 the anomalously large mass of $\eta'$ is due to the 
 mixing of the gluon state with the flavor-singlet quark
 state through the axial anomaly 
 \be
 \partial^\mu J^5_{\mu}
 =
 2i\sum_f m_f \bar{q}_f \gamma_5 q_f
 +
 2 N_f {g^2\over 16\pi^2} \, 
 {\rm Tr}\,\left( G_{\mu\nu} \tilde{G}^{\mu\nu}\right)
 \ee
 with non-vanishing $\ave{G\tilde{G}}$. 
 This implies that the $\eta'$ mesons have a large glue content. 
 It also implies that the gluon fusion process  $gg\to \eta'$ is possible.  

 In this paper, we study the generation of $\eta'$ mesons from such 
 a gluon fusion process in proton-nucleus `$(pA)$' collisions.
 In particular, we argue that measuring the $\eta'$ meson momentum
 spectrum enables us to have a direct access to the
 gluon density in the small $x$ region of the heavy nucleus.
 This is made possible by using a recently proposed
 effective vertex by Atwood and Soni \cite{Atwood:1997bn}
 between gluons and the $\eta'$ meson. 

 In \cite{Atwood:1997bn}, as a step to explain an anomalously large
 $B$ to $\eta'$ branching fraction, the authors wrote down a
 Wess-Zumino-Witten type of effective vertex:
 \be
 T_{\lambda\gamma} \delta^{ab}
 = H(p^2, q^2, P^2)\,
 \delta^{ab} \,
 \epsilon_{\mu\nu\alpha\beta}\, 
 p^\mu \,
 q^\nu \,
 (\epsilon_p^\alpha)_\lambda \,
 (\epsilon_q^\beta )_\gamma
 \label{eq:vertex}
 \ee
 roughly corresponding to an interaction Lagrangian of the form 
 ${\cal L} \sim \eta' G\tilde G$.
 Here, $p,q$ are the gluons momenta, $\epsilon_{p,q}^{\alpha,\beta}$ are the
 associated polarization vectors, and the $\delta^{ab}$ denotes the singlet
 combination of the gluons.  From now on, we denote the $\eta'$ momentum by
 a capital letter $P$.  The factor $H(p^2, q^2, P^2)$ is a form factor.
 By analyzing $J/\psi\to \eta'\gamma$ process,
 the authors of \cite{Atwood:1997bn}
 then estimate that in the on-shell limit
 \be
 H_0 \equiv H(0,0,M^2) \approx 1.8\,\GeV^{-1}
 \ee
 where $M = 0.958\GeV$ is the mass of the $\eta'$ meson.
 This vertex was then used to analyze the $B$-boson decay
 to $\eta'$ assuming that the $p^2$ and $q^2$ dependence of $H$ is
 weak\footnote{
 This weak dependence assumption has been questioned by Kagan and Petrov
 \cite{Kagan:1997qn} who
 argued that when one of the gluon momenta is off-shell,
 the form factor has to be replaced by
 \be
 H(q^2, p^2, M^2) \approx H_0 {M^2 /(M^2 - q^2)} 
 \,.
 \ee
 However, this is not relevant to the present study.
 }.
 This form of effective vertex has been also confirmed by a calculation of
 the triangle diagram contribution to the axial anomaly
 \cite{Muta:1999ue}.

 The above effective vertex represents 
 a {\em hadronization matrix element} between a two-gluon state and a hadron
 state.  To the authors' knowledge, this is unique.   
 Certainly, many constituent
 quark models can relate valence quarks to hadrons.  However, 
 except for the above $gg\eta'$ effective vertex,
 there is no other known matrix element between 
 the gluons and a known hadron state. 
 We also note that 
 the momenta involved
 in this process is not so soft compared to $\Lambda_{\rm QCD}$ or the pion
 mass since the $\eta'$ mass is almost $1\GeV$.

 Recently, these interesting features of the $gg\eta'$ vertex were exploited
 to calculate the $\eta'$ production from 
 a hadronizing quark-gluon plasma \cite{Jeon:2001cy}.
 In that study, a Boltzmann equation was used to evolve the $\eta'$ 
 density towards the hadronization time in heavy ion collisions.
 To use the $gg\eta'$ vertex in this `$AA$' environment, a certain set
 of assumptions had to be made.  These assumptions mainly involve
 the as-yet-unknown in-medium properties of the particles 
 and vertices involved in the calculation.

In this paper, we focus on production of $\eta'$ in $pA$ collisions
for two reasons. First, much is known about the structure of proton
from DIS experiments at HERA and elsewhere \cite{csdd}. Assuming that 
we understand
the parton distributions in a proton, $pA$ experiments are an excellent
tool with which to investigate parton distributions of nuclei at small $x$.
This becomes even more important in light of the fact that not much is
known about modification of parton distributions in nuclei at small
$x$ and intermediate $Q^2$ since all existing fixed target $eA$ 
experiments have limited coverage in $x$ and 
$Q^2$ \cite{arn,jmw1,jmw2,eks,bqv}. This region
of small $x$ and intermediate $Q^2 (\sim 1-10 \GeV^2)$ is the domain
of high gluon density QCD and semi-hard physics where interesting phenomena
such as saturation of gluons, etc. are expected to occur. $pA$ experiments
will give us a new venue, in addition to $eA$ and $AA$, in which to
investigate the high gluon density phase of QCD. 

Also, $pA$ collisions are much cleaner than $AA$ in the sense that
one can avoid many of 
the complications due to the in-medium effects. We do not expect to have 
a Quark Gluon Plasma (or a hot hadronic phase) created in $pA$ collisions
and can therefore avoid the difficulties associated with understanding 
the possible formation of plasma and its detailed properties.
In this work, we will
show how one can extract the nuclear gluon distribution function
at small $x$ from $pA$ experiments at RHIC assuming one knows the
gluon distribution function in a proton at not too small values  
(O($x \sim 0.1-0.01$)).

It should be noted that even though our experimental knowledge of the 
parton distribution functions in a proton is good, due mostly to HERA,
in some $pp$ induced processes such as $pp \rightarrow \gamma X$ or 
$pp \rightarrow \pi X$, it is 
quite common to introduce an ``intrinsic'' momentum $\ave{k_t^2}$ into
the standard parton distribution functions in order to fit the
experimental data \cite{ap}. For example, 
\be
xG(x,Q^2)=\int d^2 k_t \,xG(x,Q^2,k_t^2)\equiv \int d^2 k_t 
\,xG(x,Q^2)\,f(k_t^2)
\label{eq:mod}
\ee
where $f(k_t^2)$ is usually taken to be a Gaussian with width $\ave{k_t^2}$.
More explicitly,
\be
f(k_t^2)={1 \over \pi \ave{k_t^2}} 
\exp\left(-{k_t^2 \over \ave{k_t^2}}\right)
\label{eq:gauss}
\ee
so that $\int d^2k_t f(k_t^2) =1$. 
The value of $\ave{k_t^2}$ is extracted from best fits to experimental
data on dileptons, dijets, etc. and can be as large as $1-5 \GeV^2$ 
at the Tevatron.
Its value depends also on the energy of the collision and the process 
considered and is expected to be smaller for production of heavier
particles.

The origin of this ``intrinsic momentum'' is mostly initial
state radiation of soft gluons. A rigorous investigation of these soft
emissions has been carried out in some processes, such as dilepton
production, which effectively introduces this ``intrinsic'' momentum
into the standard collinear factorization based cross sections \cite{lsv}. 
In addition
to the ``intrinsic'' momentum due to the initial state radiation, there
is a genuine non-perturbative contribution which is energy and process
independent and is of order O($200-300 \MeV$) consistent with the
uncertainty principle. Phenomenologically, these 
results can also be obtained by using the modified parton distribution
functions as defined in (\ref{eq:mod}) without the resummation of the
soft gluon radiation. This is especially useful since this soft gluon
resummation has been carried out only for a limited number of processes.
Nevertheless, this resummation is usually taken to be the theoretical
justification for introduction of this ``intrinsic'' momentum into
the standard parton distribution functions. There is also the Cronin
effect in $pA$ collisions which can contribute to this intrinsic
momentum. Even though Cronin effect is not well understood in pQCD,
one can phenomenologically understand it in terms of multiple
scattering of the proton on the nucleus which causes the $p_t$
broadening of the proton \cite{plf}.

In large nuclei and at small $x$, one also has to include high 
gluon density effects. In \cite{mv1,mv2,mv3,k1,k2} it is shown 
that high gluon
densities lead to a potentially large intrinsic scale 
$Q^2_s(x) \sim A^{1/3}{1 \over x^{.2-.4}}$ in large nuclei. Most gluons
in the wave function of a nucleus in a high energy collision have momenta
of the order of this scale. In this paper we will investigate 
production of $\eta'$ in $pA$ collisions as means of extracting the
gluon distribution function of nuclei at small $x$ and study its 
dependence on the intrinsic momentum $\ave{k_t^2}$ in protons and on 
saturation scale $Q^2_s(x)$ in nuclei.

\section{$\eta'$ Production Cross Section}

In this section, we will calculate the $\eta'$ production in
$pA$ collisions and show how to extract the gluon distribution function
of the nucleus $A$. We will consider three distinct cases; first, we will
calculate the $\eta'$ production cross section, with the standard
gluon distribution function, and using the collinear factorization formalism.
Then, we will consider the case when the gluons in the nucleus $A$ 
are at small $x$ and include saturation effects in the nuclear
gluon distribution function. Finally, we will allow for ``intrinsic''
momentum in a proton and consider the case when the gluon
distribution function of the proton is also modified to take the
``intrinsic'' momentum of the proton into account.

\subsection{Collinear Factorized Cross Section}

Using collinear factorization theorems in perturbative QCD,
one can write the $\eta'$ production cross section in $pA$ collisions
as
 \be
 d\sigma^{pA\to\eta'X} = 
 \int dx_1 \,dx_2\, \, G^p(x_1,Q^2_f)\, G^A(x_2,Q^2_f)\,
 d\sigma^{gg\to\eta'}\,
 \label{eq:cfcs}
  \ee
 where $G(x_1,Q^2_f)$, $G(x_2,Q^2_f)$ are the gluon distribution 
function of the 
proton and nucleus respectively and $x_1$ and $x_2$ are the momentum
fraction of the incoming gluons while $x_L = 2P_z/\sqrt{s}$ and 
$x_{E} = 2E_P/\sqrt{s}$ are the 
momentum and energy fractions of the produced $\eta'$ respectively. 
The distribution functions, $G(x_1,Q^2_f)$ and $G(x_2,Q^2_f)$, 
depend on a factorization scale $Q^2_f$
which is shown explicitly and will be taken to be $\sim M^2_{\eta'}$
in numerical calculations. 
We also note that there are no intrinsic momenta
included in either distribution functions since it is not, strictly speaking,
allowed in collinear factorization formalism. 

The partonic cross-section for $gg \to \eta'$ is given by
 \be
 d\sigma^{gg\to\eta'}
 & = &
 {1\over 4\sqrt{(p{\cdot}q)^2}}\,
 {d^3 P\over (2\pi)^3 2E_P}\,
 \absol{T_{gg\to\eta'}}^2 \,
 (2\pi)^4\delta^4 (P - p - q)
 \label{eq:dsiggen}
 \ee
 where $p$ and $q$ are the momenta of the incoming gluons while  
$P$ is the momentum of the produced $\eta'$. The matrix element squared,
after averaging over the initial spin and color degrees of freedom, becomes
 \be
 \absol{T_{gg\to\eta'}}^2 = {1\over 64} H_0^2\, M^4_{\eta'}
 \label{eq:T2}
 \ee
Using (\ref{eq:T2}), the differential cross section becomes
 \be
 d\sigma^{gg\to\eta'} = {H_0^2 M^2_{\eta'} \over 128}
 {d^3 P\over (2\pi)^3 2E_P}\, (2\pi)^4\delta^4 (P - p - q)
 \label{eq:dsigma}
 \ee
This is the elementary partonic cross section which goes into our 
calculation of $\eta'$ production in $pA$ collisions. It should be noted 
that collinear factorization for $\eta'$ production has not been
explicitly proven but is analogous to using (rigorously proven) 
collinear factorization in production of heavy quarks, high $p_t$ jets, 
high mass dileptons, etc. and that since the natural scale involved
($\eta'$ mass) is not very large, we may have large higher order
corrections. This reflects itself in having cross sections which are 
quite sensitive to a change of factorization scale. Therefore, we
feel our results are most reliable for the ratio of cross sections
($pA$ over $pp$) rather than the absolute cross sections. 

Using Eq.~(\ref{eq:dsigma}) and resolving the delta functions gives 
 \be
 {d\sigma^{pA\to \eta'X} \over dx_L} 
 & = &
 {\pi\, H_0^2 M_{\eta^{\prime}}^2\over 64\, s\,x_E}\,
 G^p(x_+,Q^2_f)\, G^A(x_-,Q^2_f)
 \non
 & = &
 {\pi\, H_0^2 \over 64\, x_E}\,
 x_+ G^p(x_+,Q^2_f)\, x_- G^A(x_-,Q^2_f)
 \label{eq:cs}
 \ee 
 where
 \be
 x_\pm & \equiv &
 {x_E \pm x_L\over 2} = {E_P \pm P_z\over \sqrt{s}} 
 \ee
 In terms of the $\eta'$ rapidity $y$ or the longitudinal momentum fraction
 $x_L$, these are
 \be
 x_\pm
 & = &
 {M_{\eta^{\prime}}\, e^{\pm y}\over \sqrt{s}}
 =
 {\sqrt{x_L^2 + 4M_{\eta'}^2/s} \pm x_L\over 2}
 \label{eq:xpm_sol}
 \ee
 Here we take the positive $y$ (as well as $x_L$) to correspond to 
 the proton  direction.  For $x_L \gg M_{\eta'}/\sqrt{s}$, we can also write
 \be
 x_+ \approx x_L
 \ \ \ \hbox{and}\ \ \  
 x_- \approx {M_{\eta'}^2\over sx_L}
 \ee

Equation (\ref{eq:cs}) is a very simple and useful relation which 
enables us to 
extract the gluon distribution function of the nucleus {\it directly} 
(without any de-convolution, as is usually the case) from the
experimentally measured $\eta'$ production cross section. Here we
are assuming that the gluon distribution function of a proton is
known from HERA to good accuracy \cite{mrst} (to better than $10\%$) 
which is the 
case unless $x_+ \gsim 0.2$. Also, in the region $x_+ \le 10^{-4}$,
one also has a large uncertainty in determining the gluon distribution
function in a proton but we avoid this region by considering small
values of $x_-$ (recall $x_-$ is the momentum fraction of gluons in the
nucleus) since this is the region where nuclear modification of the gluon
distribution function is most pronounced and least known. 

Let us make a rough estimate of the $x$ values one can probe in a 
high energy $pA$ collision, such as those at RHIC or LHC. 
Since there are no experimental data on nuclear gluon distribution 
function at $x < 10^{-2}$ for $Q^2 > 1 \GeV$, we would like to focus on the
small $x$ (in the nucleus) kinematic region. This corresponds to
the case when $x_-$ in (\ref{eq:cs}) is small.  This in turn
corresponds to
the case when the measured $\eta'$ has a large (much larger than 
its mass or $P_t$) longitudinal momentum, i.e. in the forward
region ({\it c.f.} Eq.(\ref{eq:xpm_sol})).

At RHIC, the largest $\eta'$ rapidity in the CM frame is about 5.36.
Therefore in principle, the $x_-$ range is
\be
2.2\times 10^{-5} < x_- < 0.005
\ee
using Eq.(\ref{eq:xpm_sol}).
Since $x_+ x_- = M^2/s$, this range of $x_-$ corresponds to
\be
0.0005 <  x_+ < 1.0
\ee
For instance, 
if we detect an $\eta'$ at $y = 3.04$, that would correspond to
$x_+ = 0.10$ and $x_- = 2.3\times 10^{-4}$.
At LHC, the largest $\eta'$ rapidity in the CM frame is about 8.6. 
This implies that
\be
3.2\times 10^{-8} < x_- < 1.7\times 10^{-4}
\ee
Since $x_+ x_- = M^2/s$, this range of $x_-$ corresponds to
\be
1.7 \times 10^{-4} <  x_+ < 1.0
\ee
Again, if we fix $x_+ = 0.1$ or $y_{\eta'} = 6.35$, then $x_- = 3.0\times
10^{-7}$. 
These would be, by orders of magnitude, the smallest values of $x$ where
gluon distribution in a nucleus has ever been measured. For example, 
the smallest
values of $x$ measurable in fixed target DIS experiments at CERN and 
Fermilab is $x \sim 10^{-2}$ at similar $Q^2$ \cite{adam,amad}. 
Therefore, by measuring
the $\eta'$ production cross section at large rapidities, one can,
for the first time, determine the nuclear gluon distribution function
at very small $x$ and moderately large $Q^2$. This will help one 
understand the nature of nuclear shadowing in QCD and determine the
role of high parton density (higher twist) effects in nuclear shadowing.

\subsection{Intrinsic Momenta}

In this section, we will include the effects of intrinsic momenta
in the gluon distribution of both proton and nucleus and investigate 
the dependence
of $\eta'$ production cross section on the intrinsic momenta.
Since introducing the intrinsic momentum is done in the phenomenology
spirit as discussed earlier, we will take its width to be the 
saturation scale, $Q_s$ in nucleus. This comes about because 
at small values of $x$ in nuclei, one encounters the high gluon density
region of QCD where non-linear gluon recombination effects become
important\cite{glr,mq,ajmv1,ajmv2,jkmw,km}. The high gluon density 
introduces a new scale,
the saturation scale $Q_s(x)$, which grows with energy  
($Q^2_s(x) \gg \Lambda_{QCD}$ at small $x$) 
\cite{jklw1,jklw2,jklw3,jklw4,komi,kmw,ilm1,ilm2,film}. Most gluons 
in the wave function
of a nucleus have momenta of the order of $Q_s$ and therefore, one may
take the average (the width of the Gaussian) intrinsic momentum to
be this saturation scale $Q_s$. The value of $Q^2_s$ is estimated
to be $\sim 1-2 \GeV^2$ at RHIC. As a first approximation and to keep
our expressions simple, we will ignore the $x$ dependence of $Q^2_s$
and calculate the $\eta^{\prime}$ production cross section with
different values of $Q_s$.  

As a first step, we will include the intrinsic momentum in a nucleus and 
not in a proton since the nuclear intrinsic momentum is expected to be much
larger than that of a proton. Later, we will consider the most general case.
The momenta of the incoming gluons are now, including the intrinsic 
momentum $p_t \sim Q_s$ in the nucleus,
\be
q=(x_1\sqrt{s}/2,0_t, x_1\sqrt{s}/2)
\ee
and
\be
p=(x_2\sqrt{s}/2+p_t^2/2x_2\sqrt{s},p_t,-x_2\sqrt{s}/2+p_t^2/2x_2\sqrt{s})
\ee 
The $\eta'$ production cross section then becomes 
 \be
 d\sigma^{pA\to\eta'X} = \,
 \int dx_1 dx_2\,d^2p_t\,
 G^p(x_1,Q^2_f)\, G^A(x_2,Q^2_f, p_t)\, 
 d\sigma^{gg\to\eta'}\,
 \ee
where $ G^A(x,Q^2_f,p_t^2) = G^A(x,Q^2)\,f(p_t^2)$ and $f(p_t^2)$ 
is a Gaussian (\ref{eq:gauss}) with width of $Q^2_s$. 
 
Substituting Eq.~(\ref{eq:dsigma}) and using the $\delta$ functions
gives
\be
{d\sigma^{pA\to\eta'X} \over d^2 P_t\,dx_L} = \,
{\pi \, H_0^2 \over 64\, x_E}
x_1 G^p(x_1,Q^2_f) 
x_2 G^A(x_2,Q^2_f,P_t)\,
\ee
where
$x_1 = {M^2_{\eta'} / sx_-}$ and $x_2 = x_-$ and $P_t$ is the measured
transverse momentum of the produced $\eta'$. Therefore, measuring
the produced $\eta'$ at different rapidities ($x_L$) and transverse
momentum would determine the gluon distribution of nucleus at various
$x$ and scale $Q_f \sim M_{\eta'}$.

To get a feeling for the kinematics region where one can extract the
nuclear gluon distribution function, let us set $P_t = 1\GeV$ for RHIC.
The range of accessible $x_2$ for the nucleus is now
\be
3.3 \times 10^{-5} < x_2 < 0.007
\ \ \ (\hbox{RHIC}) 
\ee
which is still very small. 
For LHC, we have access to higher transverse moment.  
If we set $P_t = 5\,\GeV$.
The range of $x_2$ in this case is
\be
1.7 \times 10^{-7}
< x_2 < 9.3 \times 10^{-4} 
\ \ \ (\hbox{LHC}) 
\ee
Even for this large $P_t$, the accessible $x_2$ range goes 
much beyond the currently accessible region. 

Let us now consider the most general case when both the
incoming proton and nucleus have intrinsic momenta. It is still
expected that the intrinsic momentum of the proton should be
less than that of a nucleus. In principle, we could also introduce
a saturation scale for protons since at high enough energies (small $x$), 
the non-linearities of the gluonic fields which lead to the saturation scale
would become important even in a proton. However, in this work, 
we would like to restrict ourselves to not too small values of $x$ in a 
proton so that we can explore the very small $x$ region in a nucleus.
Therefore, we will ignore high gluon densities effects in a proton
and take the average intrinsic momentum in the proton as determined
in $pp$ experiments.

In the case when both proton and nucleus have transverse momenta, 
the momenta of the incoming gluons become
\be
q=(x_1\sqrt{s}/2+q_t^2/2x_1\sqrt{s} ,q_t, x_1\sqrt{s}/2-q_t^2/2x_1\sqrt{s})
\ee
and
\be
p=(x_2\sqrt{s}/2+p_t^2/2x_2\sqrt{s},p_t,-x_2\sqrt{s}/2+p_t^2/2x_2\sqrt{s}) 
\ee
The differential cross section then can be written as 
\be
d\sigma^{pA\to\eta'X} = \, \int dx_1 dx_2 \, d^2 p_t\, d^2q_t\,
G^p(x_1,Q^2_f,q_t)\,G^A(x_2,Q^2_f,p_t)\,
d\sigma^{gg\to\eta'}\,
\ee
which, upon using Eq.~(\ref{eq:dsigma}), becomes 
\be
{d\sigma^{pA\to\eta'X} \over dx_Ld^2 P_t}
&=&
{\pi \, H_0^2 \over 64\, x_E}
\int d^2 p_t \,
x_1 G^p(x_1,Q^2_f,q_t)\, 
x_2 G^A(x_2,Q^2_f,p_t)\,
{M^2\over \absol{ \hat{M}^2 - p_t^2 q_t^2/\hat{M}^2} }
\ee
where $q_t = P_t - p_t$
and 
\be
 \hat{M}^2(p_t, q_t)
 = 
 {1\over 2}
 \left( M^2 + 2 p_t{\cdot}q_t 
  +
  \sqrt{(M^2 + 2 p_t{\cdot}q_t)^2 - 4 p_t^2 q_t^2}
 \right)
 \ee
 with
 \be
 {x_1}
 = 
 {\hat{M}^2 + q_t^2 \over s x_-}
 \ \ \ \ \ 
 {x_2}
 = 
 {x_-\, \hat{M}^2 \over \hat{M}^2 + q_t^2},
 \ee
 using $x_\pm = (x_E \pm x_L)/2$.
The $p_t$ integration can be performed
numerically which would then directly relate the gluon distribution function
in a nucleus $G^A(x,Q^2_f)$ at $x$ and $Q^2_f \sim M^2_{\eta^{\prime}}$ 
to the measured rapidity and $P_t$ of the produced $\eta'$. 

However, since the $\eta'$ production cross section in $pA$ collisions 
is not experimentally known, we can conversely use the available
parameterizations of the nuclear gluon distribution function to predict
the $\eta'$ production cross section in $pA$ collisions, for example,
in those planned at RHIC. In this work, we will take this approach
and use two available parameterizations of the gluon distribution function
in nuclei due to \cite{eks} and \cite{bqv}.

\begin{figure}[htp]
\centering
\setlength{\epsfxsize=10cm}
\centerline{\epsffile{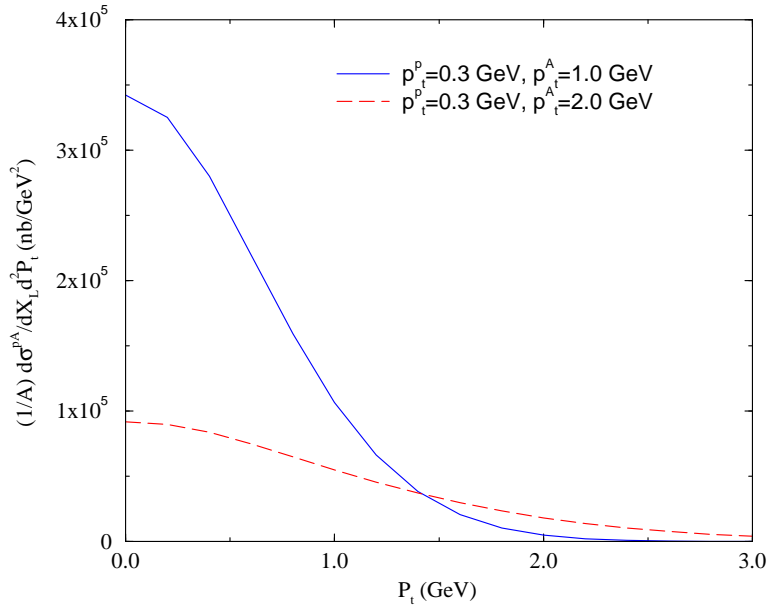}}
\caption{$\eta'$ production cross section in $pA$ at $x_L=0.1$ and 
$\sqrt{s}=200$ vs. $\eta'$ transverse momentum $P_t$.}
\label{fig:pt}
\end{figure}

In Figure (\ref{fig:pt}) we show the double differential cross section
${d\sigma^{pA\rightarrow \eta'} \over dx_Ld^2P_t}$
as a function of $\eta'$ transverse momentum at $x_L=0.1$ for two
different values of the intrinsic momentum (saturation scale) in the
nucleus. As is seen, increasing the intrinsic momentum reduces the 
differential cross section by a factor of $3-4$. It should be noted that
were we to plot our results vs. rapidity, the cross section would
increase with increasing intrinsic momentum.    

\begin{figure}[htp]
\centering
\setlength{\epsfxsize=10cm}
\centerline{\epsffile{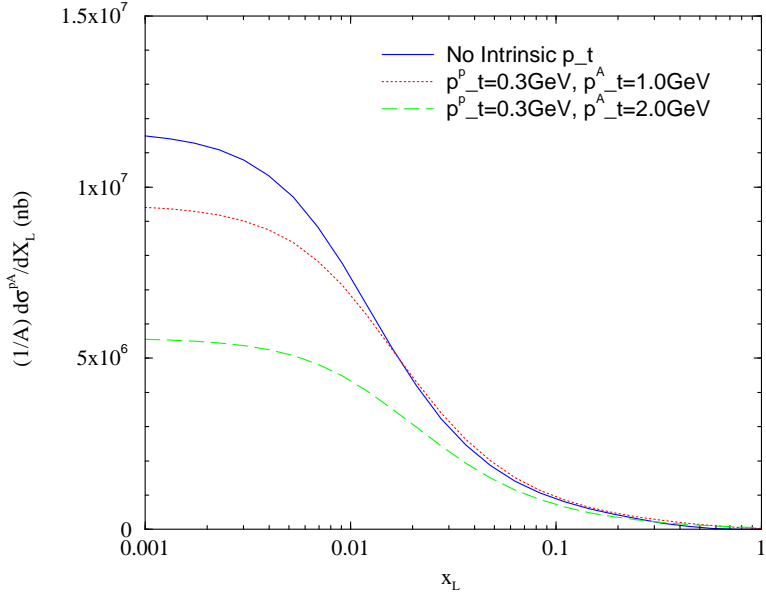}}
\caption{$\eta'$ production cross section at $\sqrt{s}=200$ vs. $x_L$.}
\label{fig:xL}
\end{figure}

In Figure (\ref{fig:xL}) we show the $P_t$ integrated cross section
vs. the $\eta'$ longitudinal momentum ratio $x_L$. Again a decrease
in the cross section is seen as we go to higher intrinsic momenta.
In both Figures (\ref{fig:pt}) and (\ref{fig:xL}) BQV shadowing~\cite{bqv}
is used.

\begin{figure}[htp]
\centering
\setlength{\epsfxsize=10cm}
\centerline{\epsffile{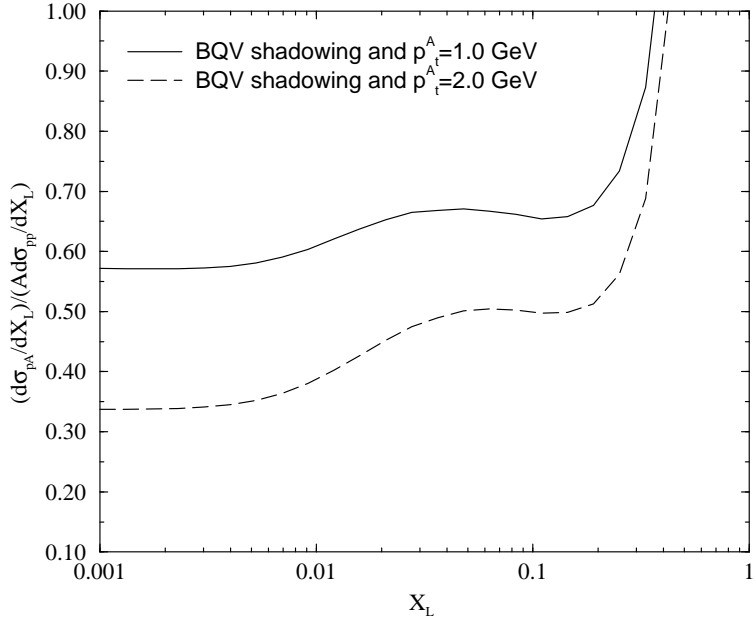}}
\caption{Ratio of $pA$ to $pp$ $\eta'$ production cross sections at 
$\sqrt{s}=200$.}
\label{fig:ratio}
\end{figure}

In Figure (\ref{fig:ratio}) we show the ratio of $\eta'$ production
cross sections in $pp$ and $pA$ collisions for two  
representative values of the nuclear intrinsic momenta. As shown, this
ratio is quite sensitive to the value of the intrinsic momenta. We
have checked that the choice of parameterization of nuclear shadowing
(\cite{eks} or \cite{bqv})  makes very little difference.

Following the thermal model estimate, we expect that 
the total cross-section for $\eta'$ production should be
about 3--5\% of $\pi^0$ cross-section.  Among the decay mode of $\eta'$, the
experimentally cleanest one is the $\eta'\to \gamma\gamma$ channel.
This process has the branching ratio of about 2\,\%. 
Therefore, the branching ratio times the cross-section should be on the
order of $A\sigma_{\pi^0}/1000 \sim 1\,$mb.
To measure this at RHIC, 
one will have to separate these photons from those
coming from other sources. Nevertheless, both STAR and PHENIX detectors
will be able to detect $\eta'$s. As a very rough estimate, PHENIX
may be able to measure $\eta'$s in the transverse momentum range
$1 < p_t < 5 $GeV and (pseudo)rapidity range $-0.3$ to $0.3$\cite{bz}.

 \section{Conclusion}

We have investigated the production of $\eta'$ in $pA$ collisions
and its sensitivity to the intrinsic momenta (saturation scale) in
nuclei. We have shown that experimental measurement of this 
cross section at RHIC would lead to determination of the gluon
distribution function in nuclei in a much wider kinematic region 
accessible by any other process in the current experiments. 
Conversely, we have used the available parameterizations of the
gluon distribution function in nuclei to predict the $\eta'$
production cross section in $pA$ collisions at RHIC.

There are several points which need to be better understood in 
order to make our calculation more accurate. First of all, we have
included only the gluon fusion and ignored all other possible 
mechanisms of $\eta'$ production. For example, it may be possible to
produce $\eta'$ through production of quarks and anti-quarks  which
would hadronize into a $\eta'$ through a completely non-perturbative
fragmentation function. The effect of this mechanism is hard to estimate
since quark (or anti-quark) fragmentation into $\eta'$ has not
been measured. 

We have also taken the $\eta'$ mass to be a constant ($958$ \MeV)
since we don't expect to have a Quark Gluon Plasma produced in a
$pp$ or $pA$ collision. One then would have had to investigate
the temperature dependence of the anomaly and include the 
poorly understood temperature dependence of the $\eta'$ mass.
However, there is a recent study of high gluon density effects on
instantons \cite{kkl} which indicates the average instanton size shrinks
as the saturation scale $Q_s$ grows analogous to the finite temperature
case. This would imply that $\eta'$ would become lighter and its
production cross section would increase at higher energies due to
the rise of $Q_s$. Decreasing $\eta'$ mass would reduce our
$gg\eta'$ matrix element (\ref{eq:T2}) and our $\eta'$ production cross 
section. Therefore measuring $\eta'$ production cross section at different
energies, for example at RHIC would help clarify this point. 

Also, we are using the collinear factorization for a process for which
there is no proof even though one may naively expect it to hold in
analogy to heavy quark production. Our factorization scale, $M^2_{\eta'}$
is not very high and our results are somewhat sensitive to the change in
this scale. Therefore, we feel that our results are most reliable
for ratio of $pA$ to $pp$ cross sections for $\eta'$ production. 

 \section*{Acknowledgement}

We would like to thank C. Gale, D. Kharzeev, L. McLerran, R. Venugopalan 
and W. Zajc for useful discussions.
J.J-M. is supported in part by a LDRD from BSA and by U.S. Department 
of Energy under Contract No. DE-AC02-98CH10886.
S.J. is supported in part by the Natural Sciences and
Engineering Research Council of Canada and by le Fonds pour la Formation
de Chercheurs et l'Aide \`a la Recherche du Qu\'ebec.

\end{document}